\documentclass[preprint,12pt,sort&compress]{elsarticle}

\usepackage{amssymb}
\usepackage{amsmath}
\usepackage{hyperref}
\usepackage{mathtools}
\usepackage{bm}

\usepackage[caption=false,labelformat=simple]{subfig}

\usepackage{pgfplots}
\pgfplotsset{compat=1.8}
\usepgfplotslibrary{statistics}
\usepackage{tikz}
\usepackage{pgfplotstable}

\pgfplotsset{
    box plot/.style={
        /pgfplots/.cd,
        black,
        only marks,
        mark=-,
        mark size=\pgfkeysvalueof{/pgfplots/box plot width},
        /pgfplots/error bars/y dir=plus,
        /pgfplots/error bars/y explicit,
        /pgfplots/table/x index=\pgfkeysvalueof{/pgfplots/box plot x index},
        line width=0.4pt
    },
    box plot box/.style={
        /pgfplots/error bars/draw error bar/.code 2 args={%
            \draw[line width=0.4pt]  ##1 -- ++(\pgfkeysvalueof{/pgfplots/box plot width},0pt) 
                |- ##2 -- ++(-\pgfkeysvalueof{/pgfplots/box plot width},0pt) |- ##1 -- cycle;
        },
        /pgfplots/table/.cd,
        y index=\pgfkeysvalueof{/pgfplots/box plot box top index},
        y error expr={
            \thisrowno{\pgfkeysvalueof{/pgfplots/box plot box bottom index}}
            - \thisrowno{\pgfkeysvalueof{/pgfplots/box plot box top index}}
        },
        /pgfplots/box plot
    },
        box plot top whisker/.style={
        /pgfplots/error bars/draw error bar/.code 2 args={%
            \pgfkeysgetvalue{/pgfplots/error bars/error mark}%
            {\pgfplotserrorbarsmark}%
            \pgfkeysgetvalue{/pgfplots/error bars/error mark options}%
            {\pgfplotserrorbarsmarkopts}%
            \draw[line width=0.4pt] ##1 -- ##2; 
        },
        /pgfplots/table/.cd,
        y index=\pgfkeysvalueof{/pgfplots/box plot whisker top index},
        y error expr={
            \thisrowno{\pgfkeysvalueof{/pgfplots/box plot box top index}}
            - \thisrowno{\pgfkeysvalueof{/pgfplots/box plot whisker top index}}
        },
        /pgfplots/box plot
    },
    box plot bottom whisker/.style={
        /pgfplots/error bars/draw error bar/.code 2 args={%
            \pgfkeysgetvalue{/pgfplots/error bars/error mark}%
            {\pgfplotserrorbarsmark}%
            \pgfkeysgetvalue{/pgfplots/error bars/error mark options}%
            {\pgfplotserrorbarsmarkopts}%
            \draw[line width=0.4pt] ##1 -- ##2; 
        },
        /pgfplots/table/.cd,
        y index=\pgfkeysvalueof{/pgfplots/box plot whisker bottom index},
        y error expr={
            \thisrowno{\pgfkeysvalueof{/pgfplots/box plot box bottom index}}
            - \thisrowno{\pgfkeysvalueof{/pgfplots/box plot whisker bottom index}}
        },
        /pgfplots/box plot
    },
    box plot median/.style={
        /pgfplots/box plot,
        /pgfplots/table/y index=\pgfkeysvalueof{/pgfplots/box plot median index},
        semithick,
        black,
        line width=0.4pt 
    },
    box plot width/.initial=1em,
    box plot x index/.initial=0,
    box plot median index/.initial=1,
    box plot box top index/.initial=2,
    box plot box bottom index/.initial=3,
    box plot whisker top index/.initial=4,
    box plot whisker bottom index/.initial=5,
}

\newcommand{\boxplot}[2][]{
    \addplot [box plot median,#1] table {#2};
    \addplot [forget plot, box plot box,#1] table {#2};
    \addplot [forget plot, box plot top whisker,#1] table {#2};
    \addplot [forget plot, box plot bottom whisker,#1] table {#2};
}

\pgfplotsset{legend image with text/.style={legend image code/.code={%
\node[anchor=west, align=right] at (0.0cm,0cm) {#1};}},}

\journal{Nuclear Physics B}

\begin{document}
\begin{frontmatter}



\title{JuliaGrid: An Open-Source Julia-Based Framework for Power System State Estimation}

\author[ETF,IVI]{Mirsad~Cosovic}
\author[IVI]{Ognjen~Kundacina}
\author[ETF]{Muhamed~Delalic}
\author[RWTH]{Armin~Teskeredzic}
\author[ETF]{Darijo~Raca}
\author[Siemens]{Amer~Mesanovic}
\author[IVI]{Dragisa~Miskovic}
\author[FTN]{Dejan Vukobratovic}
\author[RWTH,Fraunhofer]{Antonello Monti}

\affiliation[ETF]{
    organization={Faculty of Electrical Engineering, University of Sarajevo},
    country={Bosnia and Herzegovina}
}

\affiliation[IVI]{
    organization={Institute for Artificial Intelligence Research and Development},
    city={Novi Sad},
    country={Serbia}
}

\affiliation[RWTH]{
    organization={Institute for Automation of Complex Power Systems, E.ON Energy Research Center, RWTH Aachen University},
    country={Germany}
}

\affiliation[Siemens]{
    organization={Siemens AG},
    city={Munich},
    country={Germany}
}

\affiliation[FTN]{
    organization={Faculty of Technical Sciences, University of Novi Sad},
    country={Serbia}
}

\affiliation[Fraunhofer]{
    organization={Fraunhofer FIT},
    city={Aachen},
    country={Germany}
}

\begin{abstract}
Modern electric power systems have an increasingly complex structure due to rise in power demand and integration of diverse energy sources. Monitoring these large-scale systems, which relies on efficient state estimation, represents a challenging computational task and requires efficient simulation tools for power system steady-state analyses. Motivated by this observation, we propose JuliaGrid, an open-source framework written in the Julia programming language, designed for high performance execution across multiple platforms. The framework implements observability analysis, weighted least-squares and least-absolute value estimators, bad data analysis, and various algorithms related to phasor measurements. To complete power system analysis, the framework includes power flow and optimal power flow, enabling measurement generation for the state estimation routines. Leveraging computationally efficient algorithms, JuliaGrid solves large-scale systems across all methods, offering competitive performance compared to other open-source tools. It is specifically designed for quasi-steady-state analysis, with automatic detection and reuse of computed data to boost performance. These capabilities are validated on systems with 10\,000, 20\,000 and 70\,000 buses.
\end{abstract}




\end{frontmatter}



\section{Introduction}
Electric power systems are a vital fabric of modern society, underpinning technological and industrial progress. Traditionally, these systems follow a hierarchical structure comprising generation, transmission, and consumption components, with power flowing from large producers to consumers. However, the increasing integration of renewable energy sources and the deregulation of power systems have resulted in more complex control, leading to a degradation in their optimal performance. Therefore, power systems monitoring becomes a key building block to enhance control and power flow efficiency. Reliable monitoring depends on computationally efficient and accurate state estimation, which represents one of the fundamental components of modern energy management systems.

Supervisory control and data acquisition (SCADA) technology gathers input data for state estimation. SCADA provides the communication infrastructure to collect legacy measurements, including bus voltage magnitudes, bus active and reactive power injections, branch current magnitudes, and branch active and reactive power flows. Over the past few decades, the development of phasor measurement units (PMUs) enabled the measurement of bus voltage and branch current phasors with high accuracy and high sampling rates. PMUs have been instrumental in the development of wide-area measurement systems (WAMSs), which aim to provide real-time monitoring and control of power systems~\cite{gomez2011use}.

After collecting measurements at control centers, state estimation determines the present state of the power system, inherently defined by known bus voltages~\cite[Sec.1.2]{abur}. This process involves four key routines: the network topology processor, observability analysis, the state estimation algorithm, and bad data analysis. The network topology processor uses circuit breaker status information to form an energized topology, creating a bus/branch model~\cite{sasson1973automatic}. Observability analysis determines whether equations solved by the state estimation algorithm have a unique solution (i.e., whether the system is observable). If unobservable, observability analysis requires additional measurements, known as pseudo-measurements. The analysis identifies the minimal set of pseudo-measurements needed to make the system observable, ensuring a unique state estimate~\cite{monticelli1985network}. The state estimation algorithm obtains the state estimate by filtering noise-induced and redundant measurements~\cite{schweppe1970exact}. Finally, bad data analysis detects and eliminates measurement errors (i.e., outliers), followed by a new execution of the state estimation algorithm~\cite{handschin1975bad}.

\subsection{Motivation}
The scale and complexity of power systems impact state estimation, driving extensive research efforts to develop new approaches and computationally efficient algorithms for handling all state estimation routines~\cite{korresDistributed, complexBP, cosovicBP, yilmaz, muscas2015multiarea, zargar2020multiarea}. In this context, a comprehensive framework enabling comparison of novel to the state-of-the-art algorithms, represents an invaluable resource. JuliaGrid addresses these needs by providing a complete solution: from artificially generating measurements based on power flow, optimal power flow, or using real data, to determining optimal PMU placement. Moreover, JuliaGrid includes highly optimized and computationally efficient algorithms for observability analysis, state estimation algorithms, and bad data analysis, all tailored for large-scale power systems.

We host JuliaGrid on GitHub, one of the most popular platform for sharing and managing open-source code. In addition to code itself, we provide extensive documentation alongside the framework\footnote{\url{https://mcosovic.github.io/JuliaGrid.jl/stable/}}. The provided online documentation consists of manual, tutorials, and examples. The manual outlines how to use available functions, function and object signatures, and provides instructions for various analysis types. The tutorials cover theoretical underpinnings of state-of-the-art algorithms and provide detailed explanation for all mathematical models. The examples provide a wide range of toy examples coupled with various power system datasets highlighting JuliaGrid's abilities in steady-state and quasi-steady-state\footnote{The term quasi-steady-state analysis denotes a sequence of steady-state computations over consecutive time steps, and JuliaGrid accelerates this process by reusing previously computed vectors, matrices, and factorizations to efficiently evaluate slowly varying system conditions without rebuilding models from scratch.} analyses. This comprehensive documentation enhances JuliaGrid’s educational value, allowing both students and researchers to develop an in-depth understanding of mathematical models and their practical implementation in power system steady-state analyses.

\subsection{Open-Source Steady-State Power System Frameworks}
Over the past three decades, open-source power system frameworks gained significant popularity and availability, resulting in the creation of numerous advanced simulation frameworks~\cite{groissbock2019open}. One of the earliest and the most significant framework in this domain is {\scalebox{0.85}{MATPOWER}}~\cite{zimmerman2010matpower}, which has greatly influenced both education and research communities due to its well-structured and easily understandable open-source {\scalebox{0.85}{MATLAB}} code. {\scalebox{0.85}{MATPOWER}} excels in solving power flow and optimal power flow problems, and includes an optimal scheduling feature for market simulations~\cite{zimmerman2013matpower}. Despite its strengths, {\scalebox{0.85}{MATPOWER}}'s key limitation lies in its dependence on commercial software platform. The issue is addressed by {\scalebox{0.85}{PYPOWER}}~\cite{lincoln2011learning}, a Python-based adaptation of {\scalebox{0.85}{MATPOWER}}, which serves as a complete open-source alternative. More recently, pandapower adopted the {\scalebox{0.85}{PYPOWER}} framework, focusing on the automation of quasi-steady-state analysis and optimization for balanced three-phase power systems. Unlike {\scalebox{0.85}{MATPOWER}} and {\scalebox{0.85}{PYPOWER}}, pandapower allows defining power systems using nameplate parameters, further enhancing its utility and flexibility~\cite{thurner2018pandapower}. {\scalebox{0.85}{PSAT}} and PowSyBl represent frameworks based on {\scalebox{0.85}{MATLAB}}/Si\-mu\-link and Python/Java, respectively, supporting pofer flow, optimal power flow, and time-domain simulations, making it a versatile resource for power system analysis~\cite{milano2005open}.

In addition to frameworks based on {\scalebox{0.85}{MATLAB}} and Python, OpenDSS, GridLAB-D, and PowerModels represent frameworks with inherent educational and research purposes based on Delphi, C/C++, and Julia programming languages, respectively. OpenDSS offers a wide range of functionalities, primarily focused on distribution systems and distributed energy resources~\cite{dugan2011open}. GridLAB-D is an advanced tool for static simulations using agent-based models~\cite{chassin2008gridlab}. PowerModels is an open-source framework designed for solving power flow problems, leveraging just-in-time compilation to enhance performances~\cite{coffrin2018powermodels}. Finally, OpenModelica represents a general purpose framework supporting power system analyses through different third-party packages such as PowerGrids and PowerSystems~\cite{openmodelica, franke2014flexible}.

Among the primary steady-state analyses, state estimation has received comparatively less attention than power flow and optimal power flow, which have dominated most research efforts. Pandapower is one of the first to address this gap by introducing state estimation routines, including nonlinear model using legacy measurements with weighted least-squares (WLS), robust WLS, and least-absolute value (LAV) estimators, along with bad data analysis. However, pandapower lacks observability analysis, integration of PMUs in various coordinate systems, and algorithms related to PMUs, such as linear model using phasor measurements only. Additionally, some of the implemented routines may encounter issues when applied to large-scale power systems. For example, in systems with tens of thousands of buses, the bad data analysis failed due to the allocation of an extremely large amount of memory. Next, PowerGridModel, implemented in Python and C, supports state estimation for distribution power systems and observability analysis, missing the rest of the routines~\cite{PowerGridModel}. 

Recent efforts tried to address some of these missing features, with authors developing educational tools for state estimation in power systems using Python~\cite{do2022educational}. While this tool is excellent for educational purposes, it lacks several key algorithms. For example, more robust WLS methods for handling ill-conditioned cases and the implementation of a LAV estimator are missing. Additionally, there is a need for more efficient implementations of existing algorithms to support large-scale analysis, like the use of sparse inverse techniques for bad data analysis. Furthermore, the framework does not account for phase-shifting transformers, limiting its applicability to standardized power system data, such as those used in IEEE test cases. While PMUs integration into state estimation models is possible (unlike in pandapower), the framework lacks the ability to handle PMUs in different coordinate systems. 

To the best of our knowledge, JuliaGrid is the first open-source framework designed to meet these requirements by providing state-of-the-art state estimation routines, optimized for large-scale power systems. It is important to emphasize that JuliaGrid is not intended to be compared directly with commercial software such as {\scalebox{0.85}{DIgSILENT}}, PowerFactory, {\scalebox{0.85}{PSSE}}, or {\scalebox{0.85}{NEPLAN}}. Instead, its purpose is to provide an open-source alternative that is transparent, freely accessible, and tailored for research and educational use. Building on this foundation, the next subsection presents the key design principles and architecture of JuliaGrid, highlighting its educational and research-oriented objectives.

\subsection{Overview of JuliaGrid Design and Features}
JuliaGrid is implemented in the Julia programming language, utilizing just-in-time compilation to deliver computationally fast and efficient performance for steady-state and especially quasi-steady-state analysis, while also ensuring cross-platform compatibility. Among open-source tools, JuliaGrid is the only one specifically built to provide easy access to quasi-steady-state analysis and to support a wide range of power system analyses, including power flow, optimal power flow, observability analysis, state estimation, and bad data analysis. Additionally, it supports multiple unit systems for input data and offers flexible options for printing and exporting results.

The object-oriented paradigm allows the power system model, measurement data, and each analysis to be represented as objects that can be analyzed, modified, updated, or passed to different functions. More precisely, this design provides direct access to vectors, matrices, and factorized matrices at each step within the power flow and WLS state estimation algorithms. This level of control is further enhanced by support for multiple factorization methods, including LDL, LU, QR, and KLU, which is based on the Gilbert–Peierls algorithm \cite{davis2010algorithm}, enabling the solution process to be tailored to specific performance targets or application needs. Building on this foundation, JuliaGrid integrates the optimal power flow and LAV state estimation algorithms using the JuMP package~\cite{Lubin2023}, a domain-specific modeling language for mathematical optimization. Renowned in the research community for its versatility, JuMP enables JuliaGrid to run a wide range of optimization solvers (e.g., Ipopt and Gurobi) for tackling large-scale systems. To enhance flexibility, JuliaGrid facilitates the seamless extension of optimal power flow models, enabling variables and constraints to be added, updated, or removed without rebuilding the model. 

JuliaGrid core design principle resolves around support for large-scale analysis, employing computationally efficient algorithms. For example, JuliaGrid leverages Julia's column-major order for storing multidimensional arrays, by constructing matrices through accessing columns during all analyses. This strategy minimizes memory access overhead and improves cache efficiency by aligning operations with the underlying memory layout, resulting in faster data retrieval and more efficient computations.

These capabilities not only facilitate flexible and efficient computation but also enable researchers to explore complex power system analyses that were previously challenging in open-source frameworks. 
JuliaGrid is the first framework to integrate power system modeling with measurement data while enabling observability analysis, state estimation, and bad data analysis within a single runtime. This integrated and modular approach significantly lowers the barrier for testing new algorithms and analyzing large-scale systems, supporting both research innovation and educational applications.

In the following sections, beside theoretical background of state-of-the-art power system steady-state modeling and related analyses, we outline framework's key implementation design choices ensuring high computational efficiency as illustrated in performance evaluation section. 

\section{Steady-State Power System Models}
\begin{figure}[ht]
    \centering
    \includegraphics[width=8.5cm]{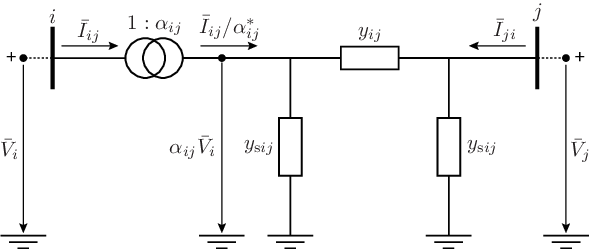}
    \caption{The unified branch model for modeling transmission lines, in-phase transformers, and phase-shifting transformers.}
    \label{fig:pi_model}
\end{figure}

A common approach to describing power system network topology is the bus/branch model, represented by a graph $\mathcal{G} = (\mathcal{N}, \mathcal{E})$. The set of nodes $\mathcal{N} = \{1, \dots, n\}$ corresponds to buses, while the set of edges $\mathcal{E} \subseteq \mathcal{N} \times \mathcal{N}$ represents the branches of the power system. Each branch is described using a two-port $\pi$-model, resulting in the unified branch model, illustrated in \figurename~\ref{fig:pi_model}. The unified branch model is characterized by its series admittance $y_{ij}$, shunt admittance $y_{\mathrm{s}ij}$, and transformer complex ratio $\alpha_{ij}$, where $(i, j) \in \mathcal{E}$.

The series admittance $y_{ij}$ is defined as the reciprocal of the series imped\-ance $z_{ij} = r_{ij} + \mathrm{j}x_{ij}$, where $r_{ij}$ represents the resistance and $x_{ij}$ is the reactance of the branch. The shunt admittances are given by $y_{\mathrm{s}ij} = g_{\mathrm{s}ij} + \mathrm{j}b_{\mathrm{s}ij}$, where $g_{\mathrm{s}ij}$ is the shunt conductance and $b_{\mathrm{s}ij}$ is the shunt susceptance. Both of these parameters take positive values for real transmission lines. The shunt conductance $g_{\mathrm{s}ij}$, often negligibly small, can be omitted in many cases. However, we retain the flexibility to include $g_{\mathrm{s}ij}$ in the model.

The transformer complex ratio is defined as $\alpha_{ij} = \tau_{ij}^{-1}e^{-\text{j}\phi_{ij}}$, where $\tau_{ij} \neq 0$ is the off-nominal transformer turns ratio, while $\phi_{ij}$ is the transformer phase shift angle. The primary side of the transformer is always considered to be connected to the $i \in \mathcal{N}$ bus of the branch $(i, j) \in \mathcal{E}$. Therefore, the admittance $y_{ij}$ is specified on the secondary side of the transformer. This distinction is important because JuliaGrid allows the definition of branch parameters not only in per-unit values but also in ohms and siemens.

In addition to these parameters, power systems may include capacitor banks or inductors for reactive power compensation. Depending on the analysis requirements, additional data, such as generator outputs and power demands at buses, can be incorporated into the model. To load power system data, JuliaGrid provides diverse input options to accommodate various requirements. Data can be constructed from scratch using built-in functions, imported from {\scalebox{0.85}{MATPOWER}} or PSSE case files, or loaded from the hierarchical data format version 5 (HDF5). The HDF5 format is an open-source standard excelling at managing large and heterogeneous datasets. Because of its efficient data handling and rapid loading capabilities, HDF5 is well-suited for large-scale power systems. JuliaGrid eases the use of this format by offering a built-in function to save constructed power system data as HDF5 file, facilitating its future reuse and supporting streamlined workflows with improved performance.

\subsection{AC Model}
We use the AC model, a fundamental component of our framework, to analyze power systems without approximations. Using Kirchhoff's circuit laws, the unified branch model can be described by the following complex expression:
\begin{equation}
  \begin{bmatrix}
    \bar{I}_{ij} \\ \bar{I}_{ji}
  \end{bmatrix} =
  \begin{bmatrix}
    \cfrac{1}{\tau_{ij}^2}({y}_{ij} + y_{\text{s}ij}) & -\alpha_{ij}^*{y}_{ij}\\
    -\alpha_{ij}{y}_{ij} & {y}_{ij} + y_{\text{s}ij}
  \end{bmatrix}
  \begin{bmatrix}
    \bar{V}_i \\ \bar{V}_j
  \end{bmatrix},
  \label{ACPiModel}
\end{equation}
where $\bar{I}_{ij}$ and $\bar{I}_{ji}$, $(i,j) \in \mathcal{E}$, represent the complex currents at the two sides of the branch, while $\bar{V}_i$ and $\bar{V}_j$ represent the complex bus voltages at buses $i \in \mathcal{N}$ and $j \in \mathcal{N}$, respectively.

As a result, derivation of all equations used in JuliaGrid related to the AC model follows from \eqref{ACPiModel}. These include equations for active and reactive power flows and injections, current magnitudes and angles in both polar and rectangular coordinate systems, as well as the real and imaginary components of the currents.

\subsection{DC Model}
The DC model is derived by linearizing the AC model defined by~\eqref{ACPiModel}. For high-voltage transmission networks under steady-state operating conditions, the voltage angle difference between adjacent buses $(i,j) \in \mathcal{E}$ is small $(\theta_{i}-\theta_{j} \approx 0)$, implying $\cos (\theta_{i}-\theta_{j})\approx 1$ and $\sin (\theta_{i}-\theta_{j}) \approx \theta_{i}-\theta_{j}$. Additionally, all bus voltage magnitudes are approximately $V_i \approx 1$, $i \in \mathcal{N}$, while branch series resistances and shunt admittances are neglected. As a result, the DC model disregards reactive powers and transmission losses, accounting only for active powers. Based on these assumptions and using \eqref{ACPiModel}, the following is obtained:
\begin{equation}
  \begin{bmatrix}
    P_{ij} \\ P_{ji}
  \end{bmatrix} = \cfrac{1}{\tau_{ij}x_{ij}}
  \begin{bmatrix*}[r]
    1 && -1\\
    -1 && 1
  \end{bmatrix*}
  \begin{bmatrix}
    \theta_{i} \\ \theta_{j}
  \end{bmatrix} + \cfrac{\phi_{ij}}{\tau_{ij}x_{ij}}
  \begin{bmatrix*}[r]
    -1 \\ 1
  \end{bmatrix*},
  \label{DCPiModel}
\end{equation}
where $P_{ij}$ and $P_{ji}$, $(i,j) \in \mathcal{E}$, represent the active power flows at the two sides of the branch, while $\theta_i$ and $\theta_j$ represent the bus voltage angles at buses $i \in \mathcal{N}$ and $j \in \mathcal{N}$, respectively. All DC analyses performed in JuliaGrid are based on the equations derived from \eqref{DCPiModel}.

\section{Power Flow and Optimal Power Flow}
The primary motivation for providing power flow and optima power flow analyses is to support state estimation as a data source for generating measurements. Furthermore, these analyses follow JuliaGrid's core design principle offering a high degree of customization. This principle allows these analyses to be used independently of the state estimation routines.

\subsection{Power Flow}
JuliaGrid supports both AC power flow and DC power flow analyses. For AC power flow, we implement solvers based on the Newton-Raphson method~\cite{tinney1967power}, the fast Newton-Raphson method with XB and BX formulations~\cite{van1989general}, and the Gauss-Seidel method~\cite{glimn1957automatic}. JuliaGrid enables straightforward updates of parameters for buses, branches, and generators, and supports the addition of new branches and generators without reconstructing the associated vectors and matrices for AC and DC models from scratch. For instance, once the nodal admittance matrices are initialized, modifications to branch parameters or status changes (e.g., from in-service to out-of-service and vice versa) require only updates to the stored matrices. These capabilities facilitate efficient execution of quasi-steady-state power flow analyses.

In addition, for AC power flow with the fast Newton-Raphson method and DC power flow, JuliaGrid tracks changes in the vectors and matrices associated with these models. The framework reuses factorized matrices whenever possible, improving computational efficiency during quasi-steady-state analyses. To illustrate, modifications to generator outputs and/or load parameters represent ideal scenario, significantly reducing the computational burden. To complement these features, JuliaGrid tracks the pattern of nonzero elements in the coefficient matrix of the DC power flow, together with the Jacobian matrices used in the Newton-Raphson and fast Newton-Raphson methods. This tracking enables in-place factorization of these matrices, reducing the number of memory allocations.

\subsection{Optimal Power Flow}
The AC and DC optimal power flow analyses in JuliaGrid are built using the JuMP package, enabling intuitive manipulation and customization of these analyses. The AC optimal power flow focuses on minimizing an objective function related to the costs of generator active and reactive power outputs, while DC optimal power flow considers only active power outputs~\cite{dommel1968optimal}. Similar to {\scalebox{0.85}{MATPOWER}}, these costs can be represented as polynomial or piecewise linear functions. For AC optimal power flow, generator capability curves can be modeled accounting for a tradeoff between the active and reactive power outputs of the generators, adding flexibility to constraint modeling. To enable a warm start of the optimal power flow, initial primal values can be manually defined or obtained from power flow results, and initial dual values can also be set.

\section{Measurement Model}
Two primary technologies, SCADA and WAMS, gather measurements across the power systems, contributing to the set of measurements $\mathcal{M}$.

Legacy measurements provided by SCADA, contributes to subsets of the set $\mathcal{M}$, each representing different types of measurements. These subsets include measurements for bus voltage magnitudes $\mathcal{V}$, branch current magnitudes $\mathcal{I}$, active power flows and injections $\mathcal{P}$, and reactive power flows and injections $\mathcal{Q}$. 

Furthermore, WAMS employs PMUs, where a single device typically measures the voltage phasor at its bus and the current phasors of all directly connected branches\footnote{Although this is not always the case. For example, some PMUs can measure current from only one incident branch.}. In the JuliaGrid framework, this concept is generalized to provide greater modeling flexibility. A PMU is represented as a set of individual phasor measurements, either bus voltage or branch current, where each phasor is defined by its magnitude, angle, and corresponding variances in the polar coordinate system. This flexible representation extends the overall measurement set $\mathcal{M}$ by introducing the subsets $\bar{\mathcal{V}}$ for bus voltage phasors and $\bar{\mathcal{I}}$ for branch current phasors. These measurements can be incorporated into state estimation models either directly in polar coordinates or, alternatively, after conversion to rectangular form. Notably, the coordinate system can be chosen independently for each individual measurement.

Each legacy measurement, as well as each magnitude and angle of phasor measurement, is associated with a measured value $z_i$, a measurement error $u_i \sim \mathcal{N} (0, v_i)$, where $v_i$ represents the variance, and a measurement function $h_i(\mathbf{x}_i)$. Here, $\mathbf{x}_i$ represents a subvector formed by selecting specific elements of the state variable vector $\mathbf{x} = [x_1, \dots, x_m]^{\mathrm{T}}$. The probability density function for the $i$-th measurement is proportional to:
\begin{equation}
  \mathcal{N}(z_i|\mathbf{x}_i,v_i) \propto \exp\Bigg\{-\cfrac{[z_i-h_i(\mathbf{x}_i)]^2}{2v_i}\Bigg\}.
  \label{gauss}
\end{equation}

In JuliaGrid the generation of measurement function $h_i(\mathbf{x}_i)$ follows the selection of state estimation model, and the function is derived from either \eqref{ACPiModel} or \eqref{DCPiModel}. The measurement value $z_i$ can be provided directly or alternatively generated artificially by adding white Gaussian noise $u_i$ to the exact value of the respective electrical quantity $e_i$, resulting in the measurement value $z_i = e_i + u_i$. 

To construct the measurement set $\mathcal{M}$, JuliaGrid provides built-in functions that use exact values obtained from one of the AC analyses. Furthermore, JuliaGrid incorporates mechanisms to randomize measurement availability, allowing specification of the number of in-service or out-of-service measurements either across the entire set or within specific subsets. Beyond these group-level operations, JuliaGrid provides capabilities for adding, updating, or modifying individual measurements, thereby facilitating faster steady-state and quasi-steady-state analyses. Once the measurement model is assembled, it can be saved to an HDF5 file for subsequent use.

Followed by definition of the measurement model based on the set $\mathcal{M}$, the relevant measurements can be selected to form the chosen state estimation model. For example, all measurements from the set $\mathcal{M}$ are included for the case of nonlinear state estimation, while the subsets of phasor measurements from $\mathcal{M}$ are used for linear state estimation with PMUs only. Therefore, depending on the selection of the state estimation model and the measurement set $\mathcal{M}$, the following system of equations can be formulated:
\begin{equation}
    \mathbf{z} = \mathbf{h}(\mathbf{x}) + \mathbf{u},
    \label{measurementModel}
\end{equation}
where $\mathbf{h}(\mathbf{x}) = [h_1(\mathbf{x}_1), \dots, h_k(\mathbf{x}_k)]^{\mathrm{T}}$ is the vector of measurement functions, $\mathbf{z} = [z_1, \dots, z_k]^{\mathrm{T}}$ is the vector of measurement values, and $\mathbf{u} = [u_1, \dots, u_k]^{\mathrm{T}}$ is the vector of measurement errors. The error vector $\mathbf{u}$ follows a zero-mean Gaussian distribution with a covariance matrix $\bm{\Sigma}$. The diagonal elements of $\bm{\Sigma}$ represent the measurement variances $\mathbf{v} = [v_1, \dots, v_k]^T$, while the off-diagonal elements correspond to the covariances between the measurement errors $\mathbf{w} = [w_1, \dots, w_k]^T$. These covariances arise only when phasor measurement functions are expressed in rectangular coordinates, which is supported by JuliaGrid. 

\section{Observability Analysis}
The state estimation algorithm aims to estimate the values of the state variables $\mathbf {x}$ based on the measurement model \eqref{measurementModel}. Prior to applying the algorithm, the observability analysis determines the existence and uniqueness of the solution for the system of equations \eqref{measurementModel}. For the cases where a unique solution is not guaranteed, the observability analysis identifies observable islands and adds set of equations defined by pseudo-measurements to achieve a unique solution~\cite{cosovic2021observability}.

In addition, JuliaGrid supports optimal PMU placement algorithms~\cite{gou2008optimal, gou2008generalized}, which identify the minimum number of PMUs required to achieve observability. In other words, this functionality constructs a measurement model \eqref{measurementModel} that guarantees a unique solution.

\subsection{Identification of Observable Islands}
JuliaGrid utilizes standard observability analysis based on the linear decoupled measurement model~\cite{monticelli1985observability}, where active power measurements from the set $\mathcal{P}$ are used to estimate bus voltage angles, while reactive power measurements from the set $\mathcal{Q}$ are used to estimate bus voltage magnitudes. This approach assumes active and reactive power measurement pairs, allowing using only active power measurements for observability analysis. Additionally, JuliaGrid provides the option to include bus voltage phasor measurements from the set $\bar{\mathcal{V}}$, enhancing the functionality for more comprehensive observability analysis.

If the system is unobservable, the observability analysis must identify all potential observable islands that can be independently solved. In this context, JuliaGrid recognizes either flow or maximal observable islands. The choice between types of islands depends on the power system's structure and the available measurements. Opting to detect only flow observable islands simplifies the island detection procedure but increases the complexity of the observability restoration, compared to identifying maximal observable islands.

To identify flow observable islands, JuliaGrid employs a topological method outlined in~\cite{horisberger1985observability}. The process begins with the examination of all active power flow measurements, aiming to determine the largest sets of connected buses linked by branches with active power flow measurements. Subsequently, the analysis considers individual boundary or tie active power injection measurements, involving two islands that may potentially be merged into a single observable island. We extend the analysis with an additional processing step to identify maximal observable islands. After processing the individual injection tie measurements, we are left with a set of injection measurements unused in the previous step. We now examine these measurements for possible merging of flow islands~\cite{horisberger1985observability}.

\subsection{Observability Restoration}
After determining the islands, the observability restoration merges these islands in a manner that protects previously determined observable states from being altered by the new set of equations defined by the pseudo-meas\-ure\-ments. This is accomplished by selecting a non-redundant subset of pseudo-meas\-ure\-ments that augment the existing set of equations with linearly independent equations~\cite[Sec. 7.3.2]{monticelliBook}.

Following the island detection step, the power system is divided into $s$ islands. Next, we focus on the set of measurements $\mathcal{M}_\mathrm{r} \subset \mathcal{P} \cup \bar{\mathcal{V}}$, which consists solely of active power injection measurements at tie buses (i.e., these measurements are called discardable or irrelevant \cite{monticelli2002handling}) and bus voltage angle measurements. These measurements are retained from the step where observable islands are identified. Based on these measurements, we introduce the matrix $\mathbf{H}_\mathrm{r} \in \mathbb{R}^{r \times s}$, where $r = |\mathcal{M}_\mathrm{r}|$. This matrix represents the coefficient matrix of a reduced network, with $s$ columns corresponding to the islands and $r$ rows corresponding to the set $\mathcal{M}_\mathrm{r}$~\cite{korres2011gram, manousakis2010observability}.

The next step involves a new measurement model containing pseudo-measurements. From this set, the restoration process uses only the following measurements: active power flows between tie buses, active power injections at tie buses, and bus voltage angles. This subset of pseudo-measurements $\mathcal{M}_\mathbf{p}$ forms the reduced coefficient matrix $\mathbf{H}_\mathbf{p} \in \mathbb{R}^{p \times s}$, where $p = |\mathcal{M}_\mathbf{p}|$. Finally, we observe the coefficient matrix defined as:
\begin{equation}
  \mathbf H = \begin{bmatrix} \mathbf H_{\text{r}} \\ \mathbf H_{\text{p}} \end{bmatrix},
\end{equation}
and forms the corresponding Gram matrix:
\begin{equation}
  \mathbf D = \mathbf H \mathbf H^T.
\end{equation}
The decomposition of $\mathbf D$ into its $\mathbf Q$ and $\mathbf R$ factors is achieved through QR factorization. Measurements needed to ensure unique state estimate, are identified by the non-zero diagonal elements corresponding to pseudo-measurements in $\mathbf{R}$. Specifically, if a diagonal element satisfies $|R_{ii}| \ge \epsilon$, JuliaGrid designates the corresponding measurement as non-redundant, where $\epsilon$ represents a pre-determined zero pivot threshold. Finally, these pseudo-measurements from the set $\mathcal{M}_\mathbf{p}$ are transferred to the set of measurements $\mathcal{M}$ for further processing, such as running the nonlinear state estimation algorithm.

\subsection{Optimal PMU Placement}
The primary goal of PMU placement algorithms is to identify the minimal set of PMUs such that the entire system is observable. In this context, two state-of-the-art approaches can be considered: optimal placement excluding and including legacy measurements. JuliaGrid implements the algorithms proposed in~\cite{gou2008optimal, gou2008generalized} to address both approaches.

The optimal PMU placement problem without considering legacy measurements is formulated as an integer linear programming problem:
\begin{equation}
  \begin{aligned}
    \text{minimize}& \;\;\; \sum_{i=1}^n d_i\\
    \text{subject\;to}& \;\;\; \sum_{j = 1}^n a_{ij}d_j \ge 1,\;\;\; \forall i \in \mathcal{N},
  \end{aligned}
\end{equation} 
where $d_i \in \mathbb{F} = \{0,1\}$ is the PMU placement decision variable associated with bus $i \in \mathcal{N}$. The binary parameter $a_{ij} \in \mathbb{F}$ indicates the connectivity of the power system network, where $a_{ij}$ can be directly derived from the nodal admittance matrix by converting its entries into binary form~\cite{xu2004observability}. This linear programming problem is implemented using JuMP allowing compatibility with different type of optimization solvers.

In the case where legacy measurements are included, JuliaGrid applies an extended formulation that incorporates existing power flow and injection measurements into the placement model. The optimal PMU placement is then obtained by solving:
\begin{equation}
  \begin{aligned}
    \text{minimize}& \;\;\; \sum_{i=1}^n d_i\\
    \text{subject\;to}& \;\;\; \sum_{j=1}^n c_{ij} \left ( \sum_{k=1}^n a_{ik}d_k \right) \geq b_i, \;\; \forall i \in \mathcal{H}.
  \end{aligned}
\end{equation}
The binary coefficients $c_{ij} \in \mathbb{F}$, $j = 1,\dots,n$, indicate the incidence of bus $i \in \mathcal{H}$ with all other buses, including itself, based on the presence or absence of measurements. In particular, the set $\mathcal{H}$ includes only buses associated with a power flow measurement, an injection measurement, or no measurement. The right-hand side coefficient $b_i$ depends on the number of nonzero terms $c_{ij}$. More precisely, three cases are distinguished. If a power flow measurement is installed at the $i$ end of branch $(i,j)$, then $c_{ii} = c_{ij} = 1$ and $b_i = 1$. The second case occurs when a power injection measurement is located at bus $i$, which is incident to buses $\mathcal{N}_i$. In this case, $c_{ii} = c_{ik} = 1$ for $k \in \mathcal{N}_i$ and $b_i = |\mathcal{N}_i|$. Finally, if bus $i$ is not incident to any measurement, then $c_{ii} = 1$ and $b_i = 1$.

From both optimization problems, the resulting binary values $d_i$ indicate where PMUs should be placed. Value $d_i = 1$ for bus $i \in \mathcal{N}$ suggests that a PMU is placed at the bus $i$. Specifically, when $d_i = 1$, a PMU is installed at bus $i$ to measure both the bus voltage phasor and all current phasors across branches incident to bus $i$.

\section{State Estimation Algorithms}
JuliaGrid implements three distinct state estimation models: nonlinear, linear using PMUs only, and a linear DC model. For each model, JuliaGrid provides an option to compute the WLS estimator using conventional methods. To address ill-conditioned scenarios, e.g., involving large discrepancies in measurement variances, JuliaGrid supports a robust WLS estimation approach based on either the orthogonal method or the Peters-Wilkinson method. Furthermore, for all three models, JuliaGrid provides the LAV estimator as a robust alternative to the WLS method, offering enhanced performance in the presence of measurement errors. 

\subsection{Nonlinear Weighted Least-Squares Estimator}
The nonlinear state estimation requires bus voltages in the polar coordinate system as state variables, which we denote by $\mathbf x \equiv [\bm {\Theta}, \mathbf{V}]^T$, where $\bm {\Theta} \in \mathbb{R}^{n-1}$ and $\mathbf {V} \in \mathbb{R}^n$ represent bus voltage angles and magnitudes, respectively. Therefore, the total number of state variables is $m = 2n-1$, accounting for the fact that the voltage angle for the slack bus is known. The nonlinear model utilizes the complete set of measurements $\mathcal{M}$, with phasor measurement functions defined in either polar or rectangular coordinates. When using rectangular coordinates, the model can include covariances or treat them as negligible.

By default, JuliaGrid represents phasor measurements in rectangular coordinates while neglecting covariances. This choice follows from the numerical robustness of the current phasor measurements representation in the rectangular coordinates. Particularly, this representation addresses ill-conditioned scenarios where current magnitudes having small values are expressed in polar coordinates~\cite{gomez2011use}.

The solution of the WLS problem, given the available set of measurements $\mathcal{M}$, can be found using the Gauss-Newton method as follows:
\begin{equation}
    \begin{gathered}
		\mathbf J (\mathbf x^{(\nu)})^{T} \bm \Sigma^{-1} \mathbf J (\mathbf x^{(\nu)}) \mathbf \Delta \mathbf x^{(\nu)} =
		\mathbf J (\mathbf x^{(\nu)})^{T} \bm \Sigma^{-1} \mathbf r (\mathbf x^{(\nu)}) \\
		\mathbf x^{(\nu+1)} = \mathbf x^{(\nu)} + \mathbf \Delta \mathbf x^{(\nu)},
        \label{gaussNewton}
    \end{gathered}
\end{equation}
where $\nu = \{0,1,2,\dots\} $ is the iteration index, $\mathbf \Delta \mathbf x \in \mathbb {R}^{m}$ is the vector of state variable increments, $\mathbf J (\mathbf x)\in \mathbb {R}^{k \times m}$ is the Jacobian matrix of measurement functions $\mathbf h (\mathbf x)$ at $\mathbf x=\mathbf x^{(\nu)}$, $\bm \Sigma \in \mathbb {R}^{k \times k}$ is a measurement error covariance matrix, and $\mathbf r (\mathbf x) = \mathbf{z} - \mathbf h (\mathbf x)$ is the vector of residuals, with $k = |\mathcal{V} \cup \mathcal{I} \cup \mathcal{P} \cup \mathcal{Q}| + 2|\bar{\mathcal{V}} \cup \bar{\mathcal{I}}|$~\cite{schweppe1970exact}.

JuliaGrid precomputes the inverse of the covariance matrix $\bm{\Sigma}$ before initiating the Gauss-Newton iterations. This approach is particularly important when covariances exist due to phasor measurements. Instead of inverting the entire covariance matrix, JuliaGrid leverages its block-diagonal structure to perform inversions of the individual block matrices, enhancing computational efficiency. 

\subsection{PMU Weighted Least-Squares Estimator}
While PMUs output phasor measurements in polar coordinates by default, these measurements can be represented in rectangular coordinates as well. For this case, the real and imaginary parts of bus voltages and branch current phasors serve as measurements. In addition, to obtain the linear system of equations, the state variables must be given in rectangular coordinates $\mathbf x \equiv[\Re{(\bar{\mathbf{V}})},\Im{(\bar{\mathbf{V}})}]^T$, where $\Re{(\bar{\mathbf{V}})} \in \mathbb {R}^{n}$ and $\Im{(\bar{\mathbf{V}})} \in \mathbb {R}^{n}$ represent the real and imaginary parts of complex bus voltages, respectively. As a result, the total number of state variables is $m = 2n$. Note that this approach does not require the slack bus~\cite{gomez2011use}.

For the case having observable power system using only PMUs, the linear WLS estimator represents the solution of the following equation:
\begin{equation}
    \mathbf H^{T} \bm \Sigma^{-1} \mathbf H \mathbf x = \mathbf H^{T} \bm \Sigma^{-1} \mathbf z,
\end{equation}
where $\mathbf z \in \mathbb {R}^{k}$ denotes the vector of measurement values, $\mathbf {H} \in \mathbb {R}^{k \times m}$ represents the coefficient matrix, and $\bm \Sigma \in \mathbb {R}^{k \times k}$ is the measurement error covariance matrix, with $k = 2|\bar{\mathcal{V}} \cup \bar{\mathcal{I}}|$.

During the transformation from polar to rectangular coordinates, measurement errors become correlated, resulting in a non-diagonal covariance matrix $\bm \Sigma$. Similar to the approach for nonlinear state estimation, JuliaGrid utilizes the block-diagonal structure of $\bm \Sigma$ to perform inversions on individual block matrices. In addition, JuliaGrid enables change to the phasor measurement variances, which only require updating the existing inverted covariance matrix. Finally, the option to neglect these correlations is also provided.

\subsection{DC Weighted Least-Squares Estimator}
Following \eqref{DCPiModel}, the DC state estimation considers only bus voltage angles, $\mathbf x \equiv \bm {\Theta} \in \mathbb{R}^{n-1}$, as the state variables~\cite{schweppe1970approximate}. As a result, the total number of state variables is $m = n-1$, where one voltage angle corresponds to the slack bus. Within the JuliaGrid framework for DC state estimation, the methodology encompasses both active power flow and injection measurements from the set $\mathcal{P}$, and bus voltage angle measurements from the set $\bar{\mathcal{V}}$. These measurements contribute to the construction of a linear system of equations \eqref{measurementModel}. 

The solution to the DC state estimation problem is determined by solving the linear WLS problem:
\begin{equation}
    \mathbf H^{T} \bm \Sigma^{-1} \mathbf H \mathbf {x} = \mathbf H^{T} \bm \Sigma^{-1} (\mathbf z - \mathbf{c}),
\end{equation}
where $\mathbf z \in \mathbb {R}^{k}$ denotes the vector of measurement values, $\mathbf {H} \in \mathbb {R}^{k \times m}$ represents the coefficient matrix, and $\bm \Sigma \in \mathbb {R}^{k \times k}$ is the diagonal covariance matrix, with $k = |\mathcal{P} \cup \bar{\mathcal{V}}|$. The inclusion of the vector $\mathbf c \in \mathbb {R}^{k}$ is necessary due to the fact that measurement functions, associated with active power measurements, may include constant terms, especially when there are non-zero shift angles of transformers or shunt elements consuming active powers.

\subsection{Orthogonal Weighted Least-Squares Method}
In general, the solution of the WLS state estimation problem using conventional methods is straightforward and efficient. However, in certain real-world systems, these methods are susceptible to numerical instabilities, preventing the algorithm from obtaining a satisfactory solution. In such cases, an alternative formulation of the WLS state estimation, known as the orthogonal method~\cite[Sec. 3.2]{abur},~\cite{costa1981robust}, can be used. To obtain the WLS estimator, JuliaGrid performs QR factorization on the rectangular matrix $\bm \Sigma^{-1/2} \mathbf J (\mathbf x^{(\nu)})$ for the nonlinear state estimation or $\bm \Sigma^{-1/2} \mathbf H$ for linear state estimation, instead of factorization of the square gain matrix (i.e., $\mathbf J (\mathbf x^{(\nu)})^{T} \bm \Sigma^{-1} \mathbf J (\mathbf x^{(\nu)})$ or $\mathbf H^{T} \bm \Sigma^{-1} \mathbf H$).  

To enhance computational efficiency and enable the method's usage in large-scale power systems, JuliaGrid avoids computational bottleneck by directly computing the matrix $\mathbf{Q}$. This approach leverages a sequence of Householder reflectors to efficiently compute the solution.

\subsection{Peters-Wilkinson Weighted Least-Squares Method}
The Peters-Wilkinson method employs LU instead of QR factorization, representing an alternative to the orthogonal method~\cite{peterswilkinson}. The solution is obtained through forward and backward substitution using the matrix $\mathbf{L}^T \mathbf{L}$ followed by backward substitution with the matrix $\mathbf{U}$ \cite[Sec.~3.4]{abur}.

\subsection{Least-Absolute Value Estimator}
The LAV method provides an alternative estimation approach, perceived as the more robust method in comparison to the methods based on the WLS~\cite{kotiuga1982lav}. The WLS state estimation methods rely on specific assumptions about measurement errors, while robust estimators aim to remain unbiased even in the presence of various types of measurement errors. This characteristic eliminates the need for bad data analysis, as discussed in~\cite[Ch. 6]{abur}. Note that robustness often comes with the cost of increased computational complexity. The LAV state estimator is derived as the solution to the optimization problem:
\begin{equation}
  \begin{aligned}
    \text{minimize}& \;\;\; \sum_{i = 1}^k |r_i|\\
    \text{subject\;to}& \;\;\; z_i - h_i(\mathbf {x}_i) =  r_i, \;\;\; i = 1, \dots, k,
  \end{aligned}
\end{equation}
where $r_i$ represents the $i$-th measurement residual. The LAV method is formulated using the JuMP package, which provides compatibility with popular optimization solvers. It is worth noting that the LAV method within JuliaGrid is available for all state estimation models.

\section{Bad Data Analysis}
Besides the WLS state estimation algorithms, bad data analysis is one of the essential routines. Its main task is to identify measurement errors, and eliminate them if possible. JuliaGrid supports bad data analysis across all implemented state estimation models.

\subsection{Chi-Squared Test}
The Chi-squared test provides a quick method to detect whether the measurement set contains outliers, without identifying which specific measurements are bad data candidates~\cite[Sec.~5.4.1]{abur}. The test is used as an initial flag before applying the more computationally intensive normalized residual test~\cite[Ch. 5]{abur}.

\subsection{Normalized Residual Test}
After computing the state estimation solution, the largest normalized residual test is iteratively applied to identify and remove bad data points one by one~\cite{korresDistributed}. Given the WLS state estimate $\hat{\mathbf{x}}$, the residuals for each measurement are calculated as:
\begin{equation}
    r_i = z_i - h_i(\hat {\mathbf {x}}_i), \;\;\; i = 1, \dots, k.
\end{equation}
The normalized residuals for all measurements are computed as follows:
\begin{equation}
    \bar{r}_{i} = \cfrac{|r_i|}{\sqrt{C_{ii}}} = \cfrac{|r_i|}{\sqrt{S_{ii}\Sigma_{ii}}}, \;\;\; i = 1, \dots, k.
\end{equation}

In this equation, we denote the diagonal entries of the residual covariance matrix $\mathbf C \in \mathbb{R}^{k \times k}$ as $C_{ii} = S_{ii}\Sigma_{ii}$, where $S_{ii}$ is the diagonal entry of the residual sensitivity matrix $\mathbf S$, representing the sensitivity of the measurement residuals to the measurement errors. For this specific configuration, the relationship is expressed as:
\begin{equation}
    \mathbf C = \mathbf S \bm \Sigma = \bm \Sigma - \mathbf J (\hat {\mathbf x}) [\mathbf J (\hat {\mathbf x})^T \bm \Sigma^{-1} \mathbf J (\hat {\mathbf x})]^{-1} \mathbf J (\hat {\mathbf x})^T.
    \label{badData}
\end{equation}
Note that only the diagonal entries of $\mathbf C$ are required. To obtain the inverse in \eqref{badData}, the JuliaGrid package utilizes a computationally efficient sparse inverse method, retrieving only the necessary elements of the inverse.

The subsequent step involves selecting the largest normalized residual that corresponds to the $j$-th measurement:
\begin{equation}
    \bar{r}_{j} = \text{max} \{\bar{r}_{i},\; i = 1, \dots, k \}.
\end{equation}
If the largest normalized residual satisfies the inequality $\bar{r}_j \ge \epsilon$, the corresponding measurement is identified as bad data and removed from the measurement set $\mathcal{M}$. Here, $\epsilon$ represents the bad data identification threshold. It is important to note that instead of simply removing the measurement, we update the existing vectors and matrices, enabling the state estimation algorithm to proceed without the need to recreate models from scratch.

\section{Performance Evaluation}
This section evaluates JuliaGrid's computational efficiency and its versatility in managing various setups under steady-state conditions. Furthermore, we test the quasi-steady-state conditions, highlighting JuliaGrid's adaptability to changes in power system and measurement models, further enhancing computational performance compared to steady-state analyses. Unlike similar open-source frameworks, JuliaGrid is explicitly designed for quasi-steady-state analysis, with internal mechanisms that automatically reuse data.

To evaluate its performance, we use power system test cases consisting of 10\,000, 25\,000, and 70\,000 buses~\cite{birchfield}, emphasizing JuliaGrid ability to handle large-scale power systems with competitive execution times\footnote{The evaluations are performed on a system running a 64-bit Ubuntu 20.04.6 LTS operating system, equipped with an AMD Ryzen 9 5950X processor (16 cores, 32 threads), 32 GB of RAM, and a 1 TB SSD.}. In all scenarios, simulations are repeated one thousand times for a given set of parameters to ensure statistically significant results. In scenarios requiring measurement sets, the measurements are randomly selected and distributed across the power system for each simulation run. Finally, we set a convergence criterion of $10^{-8}$ for all iterative algorithms.

\subsection{AC Power Flow}
To demonstrate JuliaGrid's high performance in steady-state power system analysis, we begin our evaluation with AC power flow simulations using the Newton–Raphson algorithm. Given that JuliaGrid is open-source, we restrict our benchmarks to other open-source frameworks, including {\scalebox{0.85}{MATPOWER}}, PowerModels, and pandapower. Alongside pandapower’s native AC power flow implementation, we evaluate the performance of LightSim2Grid~\cite{lightsim2grid} as a backend solver within pandapower. LightSim2Grid is a C++ solver specifically optimized for fast power flow computations\footnote{For the experiments we use JuliaGrid version 0.5.3 with Julia 1.11.2, pandapower version 3.1.2 and LightSim2Grid 0.10.3 with Python 3.12.11, and {\scalebox{0.85}{MATPOWER}} version 8.0 running on R2024a {\scalebox{0.85}{MATLAB}}.}. 

\begin{figure}[ht] 
    \centering 
    \begin{tikzpicture}
	\begin{semilogyaxis} [
            width = 13.0cm,
            height = 5.9cm,
            box plot width = 0.8mm,
    	xlabel = {Time $[\mathrm{ms}]$},
            ylabel = {Allocated Memory $[\mathrm{MiB}]$},
  	    grid = major,   		
  	    tick label style = {font=\footnotesize}, 
            label style = {font=\footnotesize},
            xticklabel style={/pgf/number format/.cd,1000 sep={}},
            ymin = 5, ymax = 2500,
            ytick={10, 100, 1000},
            legend style= {
                fill=gray!7!white, legend cell align=left, font=\tiny, at={(0.995, 0.01)}, anchor=south east,
                inner sep=2.5pt, legend columns=2, row sep=0.00cm,
  	        /tikz/every even column/.append style={column sep=0.2cm}, 
  	        /tikz/column 1/.append style={column sep=0.1cm},
                /tikz/column 2/.style={yshift=-0.025cm}, 
                /tikz/column 4/.style={yshift=-0.025cm},
                },
            nodes near coords,
            point meta = explicit symbolic,  
            every node near coord/.style={
                font=\bfseries\small,
                inner sep=0pt,
                anchor=center,
                text = black
                },
            ]   

            \pgfplotsset{juliagrid/.style={only marks,mark=*, mark size=3.5pt, mark options={fill=black!13, draw=black, line width=0.5pt}}}
            \pgfplotsset{lightsim2grid/.style={only marks, mark=square*, mark size=3.5pt, mark options={fill = black!13, draw=black, line width=0.5pt}}}
            \pgfplotsset{pandapower/.style={only marks, mark=triangle*, mark size=4.5pt, mark options={fill = black!13, draw=black, line width=0.5pt}}}
            \pgfplotsset{matpower/.style={forget plot, only marks, mark=diamond*, mark size=5.0pt, mark options={fill = black!13, draw=black, line width=0.5pt}}}
            \pgfplotsset{powermodels/.style={forget plot, only marks, mark=pentagon*, mark size=4.0pt, mark options={fill = black!13, draw=black, line width=0.5pt}}}
    
            \addlegendimage{only marks, mark=*, mark size=3.0pt, mark options={fill = black!13, draw=black, line width=0.5pt}}
            \addlegendentry{JuliaGrid}; 
            \addlegendimage{only marks}
            \addlegendentry{}

            \addlegendimage{only marks, mark=square*, mark size=3.0pt, mark options={fill = black!13, draw=black, line width=0.5pt}}
            \addlegendentry{LightSim2Grid};
            \addlegendimage{only marks, mark=-, mark size=3.0pt, mark options={line width=0.5pt}}
            \addlegendentry{10\,000-bus}; 

            \addlegendimage{only marks, mark=triangle*, mark size=3.5pt, mark options={fill = black!13, draw=black, line width=0.5pt}}
	      \addlegendentry{pandapower}; 
            \addlegendimage{only marks, mark=+, mark size=3.0pt, mark options={line width=0.5pt}}
            \addlegendentry{25\,000-bus};

            \addlegendimage{only marks, mark=diamond*, mark size=3.5pt, mark options={fill = black!13, draw=black, line width=0.5pt}}
	      \addlegendentry{{\scalebox{0.8}{MATPOWER}}}; 
            \addlegendimage{only marks, mark=star, mark size=3.0pt, mark options={line width=0.5pt}}
            \addlegendentry{70\,000-bus};

            \addlegendimage{only marks, mark=pentagon*, mark size=3.5pt, mark options={fill = black!13, draw=black, line width=0.5pt}}
            \addlegendentry{PowerModels}; 
            \addlegendimage{only marks}
            \addlegendentry{}

            \addplot[forget plot, juliagrid] coordinates {(34.376092, 27.9769516)[$-$]};
            \addplot[forget plot, juliagrid] coordinates {(99.117598, 72.12684631)[$+$]};
            \addplot[forget plot, juliagrid] coordinates {(551.259646, 209.1187668)[$\star$]};
        
            \addplot[forget plot, lightsim2grid] coordinates {(66.872358, 33.289)[$-$]};
            \addplot[forget plot, lightsim2grid] coordinates {(155.428886, 90.305)[$+$]};
            \addplot[forget plot, lightsim2grid] coordinates {(669.618487, 262.196)[$\star$]};

            \addplot[forget plot, pandapower] coordinates {(125.434279, 36.457)[\scriptsize $-$]};
            \addplot[forget plot, pandapower] coordinates {(318.529725, 134.965)[\tiny $+$]};
            \addplot[forget plot, pandapower] coordinates {(1712.77833, 292.602)[\scriptsize $\star$]};

            \addplot[forget plot, matpower] coordinates {(180.798, 38.26171875)[$-$]};
            \addplot[forget plot, matpower] coordinates {(400.59, 215.3203125)[$+$]};
            \addplot[forget plot, matpower] coordinates {(2520.5415, 1500.757813)[$\star$]};
            
            \addplot[forget plot, powermodels] coordinates {(466.22848, 261.1140366)[$-$]};
            \addplot[forget plot, powermodels] coordinates {(1680.536553, 928.7360992)[$+$]};
            \addplot[forget plot, powermodels] coordinates {(3131.020, 1381.820473)[$\star$]}; 
	\end{semilogyaxis}
    \end{tikzpicture}
    
    \caption{
        Median execution time and allocated memory (log scale) for solving the AC power flow using JuliaGrid, {\scalebox{0.85}{MATPOWER}}, PowerModels, native pandapower, and pandapower within LightSim2Grid, for power systems with 10\,000, 25\,000, and 70\,000 buses.
    }
    \label{plot:benchmark}
\end{figure}

JuliaGrid employs KLU factorization, also used by {\scalebox{0.85}{MATPOWER}} and LightSim2Grid within each iteration of the Newton–Raphson algorithm. In contrast, PowerModels relies on the external NLsolve~\cite{nlsolve} solver to handle the nonlinear system of equations, while the native pandapower AC power flow is executed using the just-in-time compiler Numba~\cite{numba}. The AC power flow is initialized using the voltage values provided in each test case. Upon convergence, we compute power flows and measure both execution time and memory allocation\footnote{Measuring memory allocation consistently across different programming languages can be challenging, particularly when functions are implemented in C or C++. The reported values represent our best effort to provide accurate and comparable results.}, as shown in \figurename~\ref{plot:benchmark}.

In general, lower execution time and reduced memory usage are preferred, both achieved by JuliaGrid. Notably, JuliaGrid successfully solves the system with 70\,000 buses in approximately the same time and memory as {\scalebox{0.85}{MATPOWER}} and PowerModels require for 25\,000-bus and 10\,000-bus system, respectively. Also, it outperforms the native pandapower AC power flow. The only tool with comparable performance is LightSim2Grid. However, even though LightSim2Grid is a C++-based implementation, JuliaGrid achieves slightly better performance. For the 70\,000-bus system, the execution time improvement is about 20\%. This result clearly demonstrates that JuliaGrid outperforms the other tools, often significantly surpassing them in both speed and memory efficiency.

\subsection{Nonlinear State Estimation}
Next, we evaluate the nonlinear WLS state estimation algorithm by only comparing the execution times of JuliaGrid and native pandapower, since the rest of frameworks do not include a built-in state estimation module. We present the normalized execution time $t_{\text{p}}/t_{\text{j}}$, where $t_{\text{p}}$ and $t_{\text{j}}$ denote execution times obtained by pandapower and JuliaGrid, respectively. For each power system, we first add a voltage magnitude measurement at each bus and active and reactive power flow measurements at each from-bus end of the branches. Then, each active and reactive power injection measurement, as well as each active and reactive power flow measurement at the to-bus end of the branches, is included independently with probability $p$. This process creates gain matrices with different densities and patterns. For instance, in a power system with 70\,000 buses, the number of non-zero elements in the gain matrix ranges from 1.4 million to 2.3 million for $p = 0.2$ and $p = 1$, respectively.
\begin{figure}[ht]
    \centering
    \begin{tikzpicture}
	\begin{axis} [
            width = 11.5cm,
            height = 5.4cm,
            box plot width = 0.8mm,
    	xlabel = {Probability that a power injection or to-bus flow measurement is included},
            ylabel = {Normalised Time $t_{\text{m}}/t_{\text{j}}$},
  	    grid = major,   		
  	    xmin = 0, xmax = 20,
            ymin = 0.6, ymax = 3.4,
  	    xtick = {1,2,3,4,5,6,7,8,9,10,11,12,13,14,15,16,17,18,19},
            xticklabels = {,$p = 0.2$,,,,$p = 0.4$,,,,$p = 0.6$,,,,$p = 0.8$,,,,$p = 1.0$}, 
  	    tick label style = {font=\footnotesize}, 
            label style = {font=\footnotesize},
  	    legend style={draw=black,fill=white,legend cell align=left,font=\tiny, at={(0.01,0.82)},anchor=west}
        ]
            \addlegendimage{area legend,fill=black!2,draw=black}
    	\addlegendentry{10\,000-bus}; 
    	\addlegendimage{area legend,fill=black!25,draw=black}
    	\addlegendentry{25\,000-bus};
            \addlegendimage{area legend,fill=black!55,draw=black}
    	\addlegendentry{70\,000-bus};
        
            \boxplot [
                forget plot, fill=black!2,
                box plot whisker bottom index = 1,
                box plot box bottom index = 2,
                box plot median index = 3,
                box plot box top index = 4,
                box plot whisker top index = 5
            ] {./boxplot/powerflow/10k.txt};
    
            \boxplot [
                forget plot, fill=black!25,
                box plot whisker bottom index = 1,
                box plot box bottom index = 2,
                box plot median index = 3,
                box plot box top index = 4,
                box plot whisker top index = 5
            ] {./boxplot/powerflow/25k.txt};
    
            \boxplot [
                forget plot, fill=black!55,
                box plot whisker bottom index = 1,
                box plot box bottom index = 2,
                box plot median index = 3,
                box plot box top index = 4,
                box plot whisker top index = 5
            ] {./boxplot/powerflow/70k.txt};
    
            \draw [black, thin] (axis cs:4,0) -- (axis cs:4,13);
            \draw [black, thin] (axis cs:8,0) -- (axis cs:8,13);
            \draw [black, thin] (axis cs:12,0) -- (axis cs:12,13);
            \draw [black, thin] (axis cs:16,0) -- (axis cs:16,13);
	\end{axis}
    \end{tikzpicture}
    
    \caption{
    Normalized execution time of the nonlinear WLS state estimation using JuliaGrid and pandapower for systems with 10\,000, 25\,000, and 70\,000 buses, shown as a function of the probability $p$ sampled independently for each active and reactive power injection, and for each active and reactive power flow at the to-bus end of the branches. Higher $p$ values yield denser gain matrices, resulting in higher computational complexity.
    }
    \label{plotacse}
\end{figure} 

\figurename~\ref{plotacse} shows that JuliaGrid outperforms pandapower across all power systems, with a clear tendency to become increasingly efficient as computational complexity grows (i.e., as $p$ increases). In these scenarios, JuliaGrid applies LU factorization in each Gauss–Newton iteration. Unlike pandapower, JuliaGrid also supports KLU factorization for state estimation, which can further enhance performance, particularly for smaller values of $p$.

\subsection{Observability Analysis, Bad Data Analysis, and PMU State Estimation}
In this subsection, we evaluate the remaining state estimation routines: observability analysis, bad data analysis, and state estimation using only PMUs. Among these, only bad data analysis is implemented in pandapower, but its implementation is limited to small systems and fails on our test cases due to excessive memory allocation. Therefore, we report execution times, shown in \figurename~\ref{plot:routine}, obtained only with JuliaGrid to demonstrate the competitive performance of these routines.

\begin{figure}[ht]
    \centering
    \begin{tikzpicture}
	\begin{semilogyaxis} [
            width = 12.5cm,
            height = 5.4cm,
            box plot width = 0.8mm,
            ylabel={Time [ms]},
  	    grid = major,   		
  	    xmin = 0, xmax = 12,
  	    xtick = {1,2,3,4,5,6,7,8,9,10,11,12},
            xticklabels = {,Observability Analysis,,,,Bad Data Analysis,,,,PMU State Estimation}, 
  	    tick label style = {font=\footnotesize}, 
            label style = {font=\footnotesize},
  	    legend style={draw=black,fill=white,legend cell align=left,font=\tiny, at={(0.01,0.82)},anchor=west}
        ]
            \addlegendimage{area legend,fill=black!2,draw=black}
    	\addlegendentry{10\,000-bus}; 
    	\addlegendimage{area legend,fill=black!25,draw=black}
    	\addlegendentry{25\,000-bus};
            \addlegendimage{area legend,fill=black!55,draw=black}
    	\addlegendentry{70\,000-bus};
        
            \boxplot [
                forget plot, fill=black!2,
                box plot whisker bottom index = 1,
                box plot box bottom index = 2,
                box plot median index = 3,
                box plot box top index = 4,
                box plot whisker top index = 5
            ] {./boxplot/routine/10k.txt};
    
            \boxplot [
                forget plot, fill=black!25,
                box plot whisker bottom index = 1,
                box plot box bottom index = 2,
                box plot median index = 3,
                box plot box top index = 4,
                box plot whisker top index = 5
            ] {./boxplot/routine/25k.txt};
    
            \boxplot [
                forget plot, fill=black!55,
                box plot whisker bottom index = 1,
                box plot box bottom index = 2,
                box plot median index = 3,
                box plot box top index = 4,
                box plot whisker top index = 5
            ] {./boxplot/routine/70k.txt};
    
            \draw [black, thin] (axis cs:4,0.1) -- (axis cs:4,3000);
            \draw [black, thin] (axis cs:8,0.1) -- (axis cs:8,3000);
	\end{semilogyaxis}
    \end{tikzpicture}
    
    \caption{Execution time (log scale) of the observability analysis including identification of maximal observable islands and the restoration step, bad data analysis, and linear WLS state estimation using PMUs only with correlated measurement errors in JuliaGrid, evaluated on power systems with 10\,000, 25\,000, and 70\,000 buses.}
    \label{plot:routine}
\end{figure} 

For the observability analysis, distinct randomly placed measurement sets are used for each power system, resulting in an average of seven maximal observable islands. The reported execution times include both the identification of maximal observable islands and the restoration process. To ensure a comprehensive evaluation, the worst-case scenario is considered by using the largest pseudo-measurement set encompassing all possible measurements. As shown, even for the power system with 70\,000 buses, JuliaGrid performs the observability analysis in approximately 100\,\text{ms}.

The bad data analysis follows after the nonlinear WLS state estimation. The reported execution times include both the detection and measurement removal processes. JuliaGrid handles the computational complexity of the normalized residual test by leveraging sparse inverse functions to retrieve only the elements necessary for identifying bad data. For example, in the 70\,000-bus power system, JuliaGrid process about 400 thousand measurements to identify the largest normalized residual value in a median time of 1343.8\,ms. We note that the obtained results strongly depend not only on the number of nonzero elements in the gain matrix, but also on their distribution pattern, since sparse inversion relies on LU factorization whose execution time is highly affected by the amount of fill-in in the $\mathbf{L}$ and $\mathbf{U}$ matrices.

Finally, we perform linear state estimation using PMUs only, with correlated measurement errors resulting in a non-diagonal covariance matrix. The number of measurement functions ranges from 100 thousand for the 10\,000-bus system to 330 thousand for the 70\,000-bus system. Despite the added complexity introduced by the non-diagonal covariance matrix, JuliaGrid demonstrates its ability to efficiently handle these analyses.

\subsection{DC Quasi-Steady-State Analysis}
Next, we evaluate JuliaGrid's efficiency in DC quasi-steady-state analysis by solving power flow and state estimation for two consecutive time steps in power systems with 10\,000 and 70\,000 buses.

\begin{figure}[ht]
    \centering 
    \captionsetup[subfigure]{oneside,margin={1.4cm,0cm}}
    \begin{tabular}{@{\hspace{-0.5cm}}c@{}} 
    \subfloat[]{\label{plot:quasidct1} \centering	
       \begin{tikzpicture}
	\begin{semilogyaxis} [
            width = 5.3cm,
            height = 5.5cm,
            box plot width = 0.8mm,
    	xlabel = {},
            ylabel={Time [ms]},
            ymode=log,
            log ticks with fixed point,
  	    grid = major,   		
  	    xmin = 0, xmax = 6,
            ymin = 0, ymax = 100,
  	    xtick = {1,2,3,4,5},
            ytick={1,3,8,22,60},
            xticklabels = {,,,}, 
            extra x ticks={1.5, 4.5},                 
            extra x tick labels={$t_1$,$t_2$},
            extra x tick style={
                tick label style={font=\footnotesize}, 
                tick style={draw=none},       
                grid=none                     
            },
  	    tick label style = {font=\footnotesize}, 
            label style = {font=\footnotesize},
  	    legend style={
                draw=black,
                fill=white,
                legend cell align=left,
                font=\tiny,
                at={(0.99,0.99)},
                anchor=north east
            }
        ]
            \addlegendimage{area legend,fill=black!2,draw=black}
    	\addlegendentry{DCPF}; 
    	\addlegendimage{area legend,fill=black!25,draw=black}
    	\addlegendentry{DCSE};
        
            \boxplot [
                forget plot, fill=black!2,
                box plot whisker bottom index = 1,
                box plot box bottom index = 2,
                box plot median index = 3,
                box plot box top index = 4,
                box plot whisker top index = 5
            ] {./boxplot/quasidc/10kpf.txt};
    
            \boxplot [
                forget plot, fill=black!25,
                box plot whisker bottom index = 1,
                box plot box bottom index = 2,
                box plot median index = 3,
                box plot box top index = 4,
                box plot whisker top index = 5
            ] {./boxplot/quasidc/10kse.txt};

            \draw [black, thin] (axis cs:3,0.1) -- (axis cs:3,200);
           
	\end{semilogyaxis}
    \end{tikzpicture}}
	\end{tabular}
	\captionsetup[subfigure]{oneside,margin={1.7cm,0cm}}
	\begin{tabular}{@{\hspace{0.3cm}}c@{\hspace{0cm}}} \subfloat[]{\label{plot:quasidct2} \centering	
         \begin{tikzpicture}
	\begin{semilogyaxis} [
            width = 5.3cm,
            height = 5.5cm,
            box plot width = 0.8mm,
    	xlabel = {},
            ylabel={Time [ms]},
            ymode=log,
            log ticks with fixed point,
  	    grid = major,   		
  	    xmin = 0, xmax = 6,
            ymin = 5, ymax = 1000,
  	    xtick = {1,2,3,4,5},
            ytick={10,27,71,189,500},
            xticklabels = {,,,}, 
            extra x ticks={1.5, 4.5},                 
            extra x tick labels={$t_1$,$t_2$},
            extra x tick style={
                tick label style={font=\footnotesize}, 
                tick style={draw=none},       
                grid=none                     
            },
  	    tick label style = {font=\footnotesize}, 
            label style = {font=\footnotesize},
  	    legend style={
                draw=black,
                fill=white,
                legend cell align=left,
                font=\tiny,
                at={(0.99,0.99)},
                anchor=north east
            }
        ]
            \addlegendimage{area legend,fill=black!2,draw=black}
    	\addlegendentry{DCPF}; 
    	\addlegendimage{area legend,fill=black!25,draw=black}
    	\addlegendentry{DCSE};
        
            \boxplot [
                forget plot, fill=black!2,
                box plot whisker bottom index = 1,
                box plot box bottom index = 2,
                box plot median index = 3,
                box plot box top index = 4,
                box plot whisker top index = 5
            ] {./boxplot/quasidc/70kpf.txt};
    
            \boxplot [
                forget plot, fill=black!25,
                box plot whisker bottom index = 1,
                box plot box bottom index = 2,
                box plot median index = 3,
                box plot box top index = 4,
                box plot whisker top index = 5
            ] {./boxplot/quasidc/70kse.txt};

            \draw [black, thin] (axis cs:3,0.1) -- (axis cs:3,1600);
           
	\end{semilogyaxis}
    \end{tikzpicture}}
	\end{tabular}
        \caption{The execution times (log scale) for quasi-steady-state analyses, including DC power flow (DCPF) and DC state estimation (DCSE), for 10\,000-bus power system (subfigure a) and 70\,000-bus power system (subfigure b) over time steps $t_1$ and $t_2$.}
	\label{plot:quasidc}
\end{figure}

First, the DC power flow model is generated and solved, where the nodal admittance matrix factorization is the primary computational task. Based on the DC power flow results, a measurement model is created, incorporating active power flows at both ends of all branches and injection measurements at all buses. This measurement model is used to formulate and solve the DC state estimation. As shown in \figurename~\ref{plot:quasidc}, the total time for $t_1$ includes the DC power flow solving time as well as the loading time for power system data from HDF5 file. For the DC state estimation, the time includes both solving and generating the measurement set using JuliaGrid's built-in functions. 

For the time step $t_2$, we randomly modify 20\% of the generator outputs and load values. JuliaGrid's built-in functions facilitate simultaneous updates to the power system data and the DC power flow model, followed by re-solving of DC power flow. The DC power flow execution time in this step is significantly lower because JuliaGrid reuses the previously factored nodal admittance matrix. Note that the reported time also includes the time required to update the power system data and the DC power flow model. For the updated power system, a new measurement set reflecting the changes is required. However, instead of creating a new measurement model, the previously defined model is updated together with the DC state estimation model. This approach reduces the execution time for the DC state estimation by reusing both the measurement model and the gain matrix factorization. JuliaGrid significantly reduces execution time for specific scenarios, particularly when the nodal admittance or gain matrix remains unchanged across time steps. 

We note that similar scenarios can be applied to the AC model, drawing similar conclusions when only loads and generator outputs are altered. In these cases, the fast Newton–Raphson method can be employed, reusing factorized matrices in each subsequent step, significantly reducing the time required to construct measurement models.

\subsection{AC Quasi-Steady-State Analysis}
Finally, JuliaGrid can reduce execution times after the initial step, even in scenarios where factorized matrices cannot be reused. The time reduction is a juxtaposition of the different approaches, such as reusing formed vectors, matrices, and using in-place factorizations. In addition, for all non-linear algorithms JuliaGrid automatically initiates a warm start for each new sequence. To illustrate, let us solve AC power flow using the Newton-Raphson method and the nonlinear state estimation using the Gauss-Newton method for power systems with 10\,000 and 70\,000 buses. As shown in \figurename~\ref{plot:quasiac} for the first time step $t_1$, the time for AC power flow includes the both loading time of the power system data with the time needed to create the AC model, and convergence time. Using results from AC power flow, we generate the measurement set and solve the nonlinear state estimation to obtain the WLS estimator, which gives the execution time for WLS state estimation. Note that the number of measurements is 32\,283 and 224\,539 for power systems with 10\,000 and 70\,000 buses, respectively. 

\begin{figure}[ht]
    \centering 
    \captionsetup[subfigure]{oneside,margin={1.6cm,0cm}}
    \begin{tabular}{@{\hspace{-0.5cm}}c@{}} 
    \subfloat[]{\label{plot:quasiact1} \centering	
       \begin{tikzpicture}
	\begin{semilogyaxis} [
            width = 5.3cm,
            height = 5.5cm,
            box plot width = 0.8mm,
    	xlabel = {},
            ylabel={Time [ms]},
            ymode=log,
            log ticks with fixed point,
  	    grid = major,   		
  	    xmin = 0, xmax = 6,
            ymin = 15, ymax = 350,
  	    xtick = {1,2,3,4,5},
            ytick={20,36,63,112,200},
            xticklabels = {,,,}, 
            extra x ticks={1.5, 4.5},                 
            extra x tick labels={$t_1$,$t_2$},
            extra x tick style={
                tick label style={font=\footnotesize}, 
                tick style={draw=none},       
                grid=none                     
            },
  	    tick label style = {font=\footnotesize}, 
            label style = {font=\footnotesize},
  	    legend style={
                draw=black,
                fill=white,
                legend cell align=left,
                font=\tiny,
                at={(0.99,0.99)},
                anchor=north east
            }
        ]
            \addlegendimage{area legend,fill=black!2,draw=black}
    	\addlegendentry{ACPF}; 
    	\addlegendimage{area legend,fill=black!25,draw=black}
    	\addlegendentry{ACSE};
        
            \boxplot [
                forget plot, fill=black!2,
                box plot whisker bottom index = 1,
                box plot box bottom index = 2,
                box plot median index = 3,
                box plot box top index = 4,
                box plot whisker top index = 5
            ] {./boxplot/quasiac/10kpf.txt};
    
            \boxplot [
                forget plot, fill=black!25,
                box plot whisker bottom index = 1,
                box plot box bottom index = 2,
                box plot median index = 3,
                box plot box top index = 4,
                box plot whisker top index = 5
            ] {./boxplot/quasiac/10kse.txt};

            \draw [black, thin] (axis cs:3,0.1) -- (axis cs:3,380);
           
	\end{semilogyaxis}
    \end{tikzpicture}}
	\end{tabular}
	\captionsetup[subfigure]{oneside,margin={1.8cm,0cm}}
	\begin{tabular}{@{\hspace{0.3cm}}c@{\hspace{0cm}}} \subfloat[]{\label{plot:quasiact2} \centering	
         \begin{tikzpicture}
	\begin{semilogyaxis} [
            width = 5.3cm,
            height = 5.5cm,
            box plot width = 0.8mm,
    	xlabel = {},
            ylabel={Time [ms]},
            ymode=log,
            log ticks with fixed point,
            yticklabel style={/pgf/number format/.cd,1000 sep={}},
  	    grid = major,   		
  	    xmin = 0, xmax = 6,
            ymin = 160, ymax = 1500,
  	    xtick = {1,2,3,4,5},
            ytick={200,300,450,670,1000},
            xticklabels = {,,,}, 
            extra x ticks={1.5, 4.5},                 
            extra x tick labels={$t_1$,$t_2$},
            extra x tick style={
                tick label style={font=\footnotesize}, 
                tick style={draw=none},       
                grid=none                     
            },
  	    tick label style = {font=\footnotesize}, 
            label style = {font=\footnotesize},
  	    legend style={
                draw=black,
                fill=white,
                legend cell align=left,
                font=\tiny,
                at={(0.99,0.99)},
                anchor=north east
            }
        ]
            \addlegendimage{area legend,fill=black!2,draw=black}
    	\addlegendentry{ACPF}; 
    	\addlegendimage{area legend,fill=black!25,draw=black}
    	\addlegendentry{ACSE};
        
            \boxplot [
                forget plot, fill=black!2,
                box plot whisker bottom index = 1,
                box plot box bottom index = 2,
                box plot median index = 3,
                box plot box top index = 4,
                box plot whisker top index = 5
            ] {./boxplot/quasiac/70kpf.txt};
    
            \boxplot [
                forget plot, fill=black!25,
                box plot whisker bottom index = 1,
                box plot box bottom index = 2,
                box plot median index = 3,
                box plot box top index = 4,
                box plot whisker top index = 5
            ] {./boxplot/quasiac/70kse.txt};

            \draw [black, thin] (axis cs:3,0.1) -- (axis cs:3,1600);
           
	\end{semilogyaxis}
    \end{tikzpicture}}
	\end{tabular}
        \caption{The execution times (log scale) for quasi-steady-state analyses, including AC power flow (ACPF) and nonlinear state estimation (ACSE), for 10\,000-bus power system (subfigure a) and 70\,000-bus power system (subfigure b) over time steps $t_1$ and $t_2$.}
	\label{plot:quasiac}
\end{figure} 

Next, we modify the off-nominal turn ratio of all transformers in the system by ±1\% followed by re-solving the AC power flow. For this modification, we update both the power system data and the AC power flow model using built-in functions. This update allows us to reuse the AC model, the Jacobian matrix pattern, and run the Newton-Raphson method in a warm start, leading to a reduced execution time for time step $t_2$. We update the previous measurement model, along with the nonlinear WLS state estimation model. This results in reusing the patterns of the measurement vector, Jacobian matrix, and the covariance matrix. Additionally, the Gauss-Newton method is automatically initialized in a warm start. This synergy of techniques results in faster execution time for the step $t_2$.

\section{Conclusions}
This paper presents the JuliaGrid, an open-source framework implemented in the Julia programming language. The majority of the JuliaGrid's codebase represents the independent implementation, minimizing reliance on external packages beyond Julia’s standard library. This approach ensures long-term support for algorithms integrated within the framework. JuliaGrid is optimized for solving large-scale problems in power flow, optimal power flow, observability analysis, state estimation algorithms, and bad data analysis. JuliaGrid’s flexibility, scalability, and quasi-steady-state capabilities open new avenues for addressing complex research questions and conducting large-scale system analyses that were previously intractable. In this way, JuliaGrid not only advances computational efficiency but also enables researchers to explore challenges such as grid operation, integration of advanced control strategies, and testing of novel algorithms in a transparent environment. Its stable architecture allows easy integration of new routines and analyses in future versions of the framework, including but not limited to advanced observability analyses, robust state estimation algorithms, and alternative bad data processing techniques, ensuring it evolves with research and educational needs.

\bibliographystyle{elsarticle-num} 
\bibliography{cas-refs.bib}

\end{document}